\documentclass[nofootinbib,aps,amsmath,amssymb,a4paper,prl,twocolumn]{revtex4-1}
\usepackage[english,american]{babel}
\usepackage[utf8]{inputenc}
\usepackage[T1]{fontenc}

\usepackage{graphicx}
\usepackage{color}

\usepackage{units}
\usepackage{hyperref}
\hypersetup{
   pdfnewwindow=true,     % links in new window
   colorlinks=true,       % false: boxed links; true: colored links
   linkcolor=black,       % color of internal links
   citecolor=blue,        % color of links to bibliography
   filecolor=black,       % color of file links
   urlcolor=blue          % color of external links
}

\begin{document}

%%%%%%%%%%%%%%%%%%%%%%%%%%%%%%%%%%%%%%%%%%%%%%%%%%%%%%%%%%%%%%%%%%%%%%%%%%%%%%%
% Title start
%%%%%%%%%%%%%%%%%%%%%%%%%%%%%%%%%%%%%%%%%%%%%%%%%%%%%%%%%%%%%%%%%%%%%%%%%%%%%%%
\title{\Large {\bf{Scalar Singlet Dark Matter and Gamma Lines}}}
\author{Michael \surname{Duerr}}
%\email{michael.duerr@mpi-hd.mpg.de}
\author{Pavel \surname{Fileviez P\'erez}}
%\email{fileviez@mpi-hd.mpg.de}
\author{Juri \surname{Smirnov}}
%\email{fileviez@mpi-hd.mpg.de}
\affiliation{\\ \small{
Particle and Astro-Particle Physics Division \\
Max-Planck-Institut f\"ur Kernphysik \\
Saupfercheckweg 1, 69117 Heidelberg, Germany}
}

%%%%%%%%%%%%%%%%%%%%%%%%%%%%%%%%%%%%%%%%%%%%%%%%%%%%%%%%%%%%%%%%%%%%%%%%%%%%%%%
%%%%%%%%%%%%%%%%%%%%%%%%%%%%%%%%%%%%%%%%%%%%%%%%%%%%%%%%%%%%%%%%%%%%%%%%%%%%%%%
\begin{abstract}
We point out the possibility to test the simplest scalar dark matter model at gamma-ray telescopes. 
We discuss the relevant constraints and show the predictions for direct detection, gamma line searches and LHC searches.   
Since the final state radiation processes are suppressed by small Yukawa couplings one could observe the gamma lines from dark matter annihilation.
\end{abstract}
%%%%%%%%%%%%%%%%%%%%%%%%%%%%%%%%%%%%%%%%%%%%%%%%%%%%%%%%%%%%%%%%%%%%%%%%%%%%%%%
%%%%%%%%%%%%%%%%%%%%%%%%%%%%%%%%%%%%%%%%%%%%%%%%%%%%%%%%%%%%%%%%%%%%%%%%%%%%%%%

\maketitle

%%%%%%%%%%%%%%%%%%%%%%%%%%%%%%%%%%%%%%%%%%%%%%%%%%%%%%%%%%%%%%%%%%%%%%%%%%%%%%%
\section{Introduction}
%%%%%%%%%%%%%%%%%%%%%%%%%%%%%%%%%%%%%%%%%%%%%%%%%%%%%%%%%%%%%%%%%%%%%%%%%%%%%%%
The possibility to describe the properties of the dark matter~(DM) in the Universe using a particle in models for physics beyond the Standard Model~(SM) of particle physics has called the 
full attention of the high-energy physics community. Currently, we know basically nothing about 
the dark matter in the Universe apart from that the DM relic density should be $\Omega_\text{DM} h^2=0.12$. We have, however, a series of important constraints coming from many types of experiments. 
Direct detection experiments constrain the interactions with the Standard Model fermions.
One important ``smoking gun'' for any dark matter model would be the existence of at least one gamma line from dark matter annihilation 
which can be used to predict the dark matter mass.

The testability of a dark matter model is a complex task. In this Letter we revisit the simplest scalar dark matter model~\cite{Silveira:1985rk} where the Standard Model is extended by a real scalar field. In this dark matter model, one can 
relate the predictions for direct and indirect experiments once we use the relic density constraints.
Therefore, one can hope to test this model at different experiments. See Refs.~\cite{McDonald:1993ex,Burgess:2000yq,O'Connell:2006wi,Barger:2007im,He:2008qm,Farina:2009ez,Guo:2010hq,Profumo:2010kp,Djouadi:2012zc,Cheung:2012xb,Cline:2013gha,Feng:2014vea,Han:2015hda,Kadastik:2011aa,Khan:2014kba,Yaguna:2008hd,Kahlhoefer:2015jma} for previous studies of this model. 

In this Letter we investigate the annihilation into gamma rays in the scalar singlet model, and our main aim is to investigate the visibility of the gamma lines. 
In this model one can have only two gamma lines coming from the annihilation into two photons and into $Z \gamma$.
In this context the dark matter 
annihilation into two Standard Model fermions and a photon is suppressed by the small bottom Yukawa coupling. This 
is the so-called final state radiation channel which typically spoils the visibility of the gamma line. Since these 
processes are suppressed one can hope to observe clearly the gamma lines at gamma ray telescopes such as 
Fermi-LAT. 

Our main result is that in the simplest scalar dark matter model one can observe the 
gamma lines when one has a good energy resolution. This is a striking result which has been overlooked in past studies 
and is crucial to test this model at gamma ray experiments. 

%%%%%%%%%%%%%%%%%%%%%%%%%%%%%%%%%%%%%%%%%%%%%%%%%%%%%%%%%%%%%%%%%%%%%%%%%%%%%%%
\section{Scalar Singlet Dark Matter}
%%%%%%%%%%%%%%%%%%%%%%%%%%%%%%%%%%%%%%%%%%%%%%%%%%%%%%%%%%%%%%%%%%%%%%%%%%%%%%%
In the simplest scalar dark matter model~\cite{Silveira:1985rk} the Standard Model is extended by one real scalar singlet $S$, and the relevant part of the Lagrangian reads as
\begin{equation}
- {\cal L} \supset \frac{1}{2} m_S^2 S^2 + \lambda_S S^4 + \lambda_p H^\dagger H S^2,
\end{equation} 
where $H$ is the SM Higgs. After electroweak symmetry breaking the mass of the scalar dark matter is given by $M_S^2=m_S^2+ \lambda_p v_0^2$, where $v_0 = \unit[246]{GeV}$ is the SM Higgs vacuum expectation value. In order to guarantee the dark matter stability a discrete ${\cal Z}_2$ symmetry is imposed under which only $S$ is odd, i.e., $S \to - S$.  

This model has only one extra degree of freedom and only two parameters, the portal coupling $\lambda_p$ and the physical dark matter mass $M_S$. The quartic coupling $\lambda_S$ does not play any role in DM phenomenology. Therefore, 
using the bounds from relic density, direct detection and indirect detection one can show in a simple way the allowed parameter space in agreement with all experiments.
See Refs.~\cite{McDonald:1993ex,Burgess:2000yq,O'Connell:2006wi,Barger:2007im,He:2008qm,Farina:2009ez,Guo:2010hq,Profumo:2010kp,Djouadi:2012zc,Cheung:2012xb,Cline:2013gha,Feng:2014vea,Han:2015hda,Kadastik:2011aa,Khan:2014kba,Yaguna:2008hd,Kahlhoefer:2015jma}
for the study of phenomenological and cosmological aspects of this model. 

\begin{figure}[t]
 \includegraphics[width=0.85\linewidth]{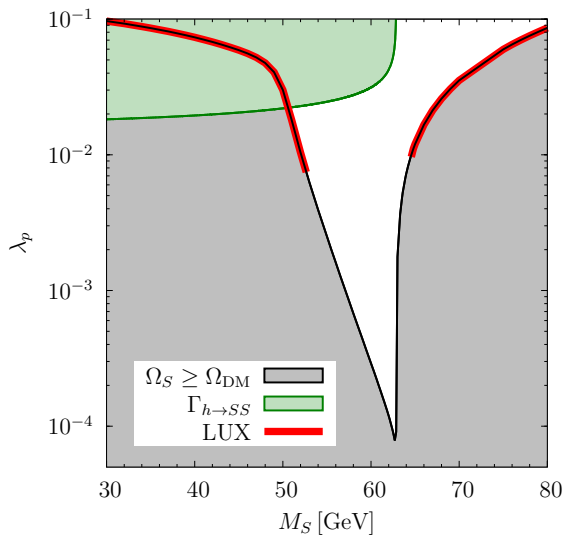}
 \caption{Allowed parameter space in the $M_S$--$\lambda_p$ plane. 
 The black line corresponds to values in agreement with the full relic density today, $\Omega_\text{DM} h^2=0.1199 \pm 0.0027$~\cite{Ade:2013zuv}, and the gray area is robustly excluded by a too large DM relic density. 
 The green area is ruled out by the invisible decay of the SM Higgs, ${\rm{BR}} (h \to SS) < 58 \%$~\cite{Chatrchyan:2014tja}.
 The lines in red correspond to the values in disagreement with the LUX results~\cite{Akerib:2013tjd}.} 
 \label{fig:parameterspace}
\end{figure}

\begin{figure}[t]
 \includegraphics[width=0.85\linewidth]{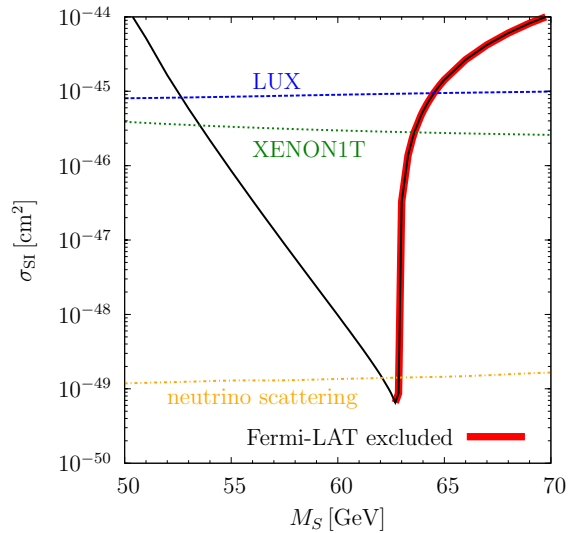}
 \caption{Spin-independent nucleon--DM cross section as a function of the dark matter mass $M_S$ assuming that $S$ makes up the full DM relic density. The LUX~\cite{Akerib:2013tjd} (blue dashed) and projected XENON1T~\cite{Aprile:2012zx} (green dotted) bounds cut into the parameter space. The red part of the curve is excluded by $b \bar{b}$ limits from Fermi-LAT~\cite{Ackermann:2015zua}, see also Fig.~\ref{fig:indirectdetection}. We show the neutrino coherent scattering background (orange dash-dotted)~\cite{Billard:2013qya}.}
 \label{fig:directdetection}
\end{figure}

To discuss the testability of this model, we focus on the low mass region ($M_S < M_h/2$) where one could hope to test the model at the LHC or at future colliders through the invisible SM Higgs decay. For $M_S > M_h/2$, it is very challenging 
to test this model at colliders; see Ref.~\cite{Craig:2014lda} for a recent discussion. 

In this low mass region, the main annihilation channels of $S$ are into two SM fermions $f$, for which the cross section is given by
\begin{equation}
\sigma ( SS \to \bar{f} f ) = \frac{\lambda_p^2 M_f^2 (s-4 M_f^2)^{3/2}}{2 \pi s \sqrt{s-4 M_S^2} \left[ (s-M_h^2)^2 + M_h^2 \Gamma_h^2 \right] }.
\end{equation}
Below threshold, the total decay width of the SM Higgs includes the invisible decay to $S$, 
\begin{equation}
 \Gamma_h = \Gamma_h^\text{SM} + \Gamma (h \to SS),
\end{equation}
where
\begin{equation}
\Gamma (h \to SS) = \frac{\lambda_p^2 v_0^2}{8 \pi M_h^2} (M_h^2-4 M_S^2)^{1/2}.
\end{equation}

Fig.~\ref{fig:parameterspace} shows the parameter space allowed by all relevant constraints. The black line corresponds to the values 
in agreement with the full relic density, $\Omega_\text{DM} h^2=0.1199 \pm 0.0027$~\cite{Ade:2013zuv}. The gray area is ruled out since in this part of the parameter space one overcloses the Universe. The green area is ruled out by the invisible decay of the SM Higgs imposing that  $\Gamma (h \to SS) < 58 \%$~\cite{Chatrchyan:2014tja}. The lines in red correspond to the values in disagreement with the LUX direct detection experiment, see below for more details.\footnote{Notice that the bound on the direct detection cross section assumes that the DM candidate under consideration makes up the full DM relic density.}
Therefore, the only allowed values for the mass $M_S$ and the coupling $\lambda_P$ are defined by the black line except for the red parts. These results are in agreement with the results presented in Ref.~\cite{Cline:2013gha}.

Using the allowed values for $M_S$ and $\lambda_P$ by relic density, we show in Fig.~\ref{fig:directdetection} the predicted values for the spin-independent nucleon--DM cross section which is given by
\begin{equation}
\sigma_{SI}=\frac{\lambda_p^2 f_N^2 \mu^2 m_N^2}{\pi M_h^4 M_S^2},
\end{equation}
where $\mu= m_N M_S/ (m_N + M_S)$. Here $m_N$ is the 
nucleon mass and $f_N=0.3$ is the matrix element~\cite{Cline:2013gha}. The blue dashed line corresponds to the current LUX upper bound on the cross section~\cite{Akerib:2013tjd}, and the projected XENON1T bound~\cite{Aprile:2012zx} is represented by the green dotted line.

\begin{figure}[t]
 \includegraphics[width=0.85\linewidth]{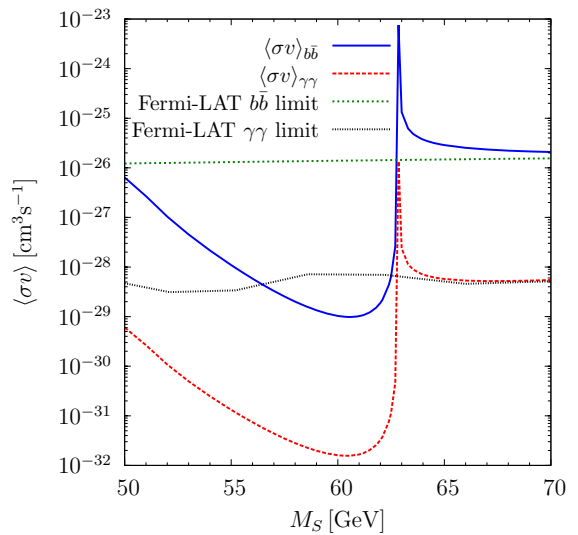}
 \caption{Velocity-averaged cross sections times velocity for the annihilation into $\bar{b} b$ (blue solid) and $\gamma \gamma$ (red dashed), and the corresponding upper bounds from Fermi-LAT~\cite{Ackermann:2015zua,Ackermann:2015lka} (green and black dotted). }
 \label{fig:indirectdetection}
\end{figure}

Knowing the results from the relic density and the constraints from direct detection experiments we are ready to study the predictions for indirect detection.
In Fig.~\ref{fig:indirectdetection} we show the predictions for the annihilation into $b\bar{b}$ (blue solid curve) as well as the current corresponding bound from Fermi-LAT~\cite{Ackermann:2015zua} (green dotted curve). Notice that the Fermi-LAT $\bar{b}b$-limit rules out the region for $M_S > 62.8$ GeV.  Therefore, only the range $\unit[53]{GeV} \leq M_S \leq \unit[62.8]{GeV}$ is allowed by the relic density, direct detection and $b\bar{b}$ constraints. This information is crucial to understand the predictions for the annihilation into gamma rays in this model, which we study in detail in the next section.

%%%%%%%%%%%%%%%%%%%%%%%%%%%%%%%%%%%%%%%%%%%%%%%%%%%%%%%%%%%%%%%%%%%%%%%%%%%%%%%
\section{Gamma Lines}
%%%%%%%%%%%%%%%%%%%%%%%%%%%%%%%%%%%%%%%%%%%%%%%%%%%%%%%%%%%%%%%%%%%%%%%%%%%%%%%
In this model one can have two possible gamma lines from dark matter annihilation, $$SS \to \gamma \gamma, Z \gamma.$$
In order to investigate the visibility of these gamma lines one has to understand the correlation between the predictions for final state radiation 
and the annihilation into gamma lines. The final state radiation in the low mass window is suppressed by the small Yukawa coupling of the bottom quark, which is the dominant contribution. To the annihilation into gamma lines, however, all charged SM fields contribute. Therefore, one can observe easily the $\gamma \gamma$ line in this scenario.

\begin{figure}[t]
 \includegraphics[width=0.9\linewidth]{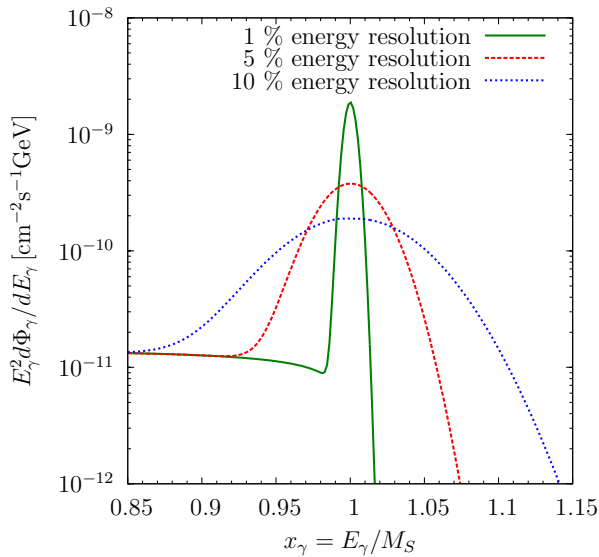}
 \caption{Spectrum of the gamma line from $SS \rightarrow \gamma \gamma$ for $1\%$ (green solid), $5\%$ (red dashed), and $10 \%$ (blue dotted) energy resolution. We use $M_S = \unit[62.5]{GeV}$ and $\lambda_p = 9.1\times 10^{-5}$, in agreement with today's relic density. For these values of the parameters, the $Z\gamma$ line is at $x_\gamma = 0.47$. For the plot, the $J$-factor for the R3 region-of-interest is used, given by the Fermi-LAT collaboration to be $J_\text{ann} = \unit[13.9\times 10^{22}]{GeV^2 cm^{-5}}$~\cite{Ackermann:2013uma}. 
 \label{fig:spectrum}} 
\end{figure}

In Fig.~\ref{fig:indirectdetection} we show the velocity-averaged cross section times velocity for the dark matter annihilation into two gammas, and corresponding limit from Fermi-LAT~\cite{Ackermann:2015lka}. The $Z \gamma$ line is at $E_\gamma \approx \unit[29.2]{GeV}$ when $M_S = \unit[62.5]{GeV}$ (the benchmark scenario we use later), and is therefore more difficult to observe due to secondary photons from pion decays. However, the visibility of the $\gamma \gamma$ line is unobstructed. 
As in the case of the annihilation into $b \bar{b}$, the region above $62.8$ GeV is ruled out. 
Notice that the region around the resonance cannot be tested in direct searches but can 
be tested in indirect detection experiments. This means that the complementarity of these experiments 
is crucial to test or rule out this model.

In Fig.~\ref{fig:spectrum} we show the predictions for the final state radiation and the gamma line assuming a Gaussian distribution 
for the detector resolution, for $1\%$ (green solid), $5\%$ (red dashed), and
$10 \%$ (blue dotted) energy resolution. We use the benchmark scenario $M_S = \unit[62.5]{GeV}$ and $\lambda_p = 9.1\times 10^{-5}$, in agreement with today's relic density.
Notice that for an energy resolution of $1 \%$ one can have a gamma line  
two orders of magnitude stronger than the final state radiation. Therefore, the gamma line can be easily observed in this model.
 
\begin{figure}[t]
 \includegraphics[width=0.85\linewidth]{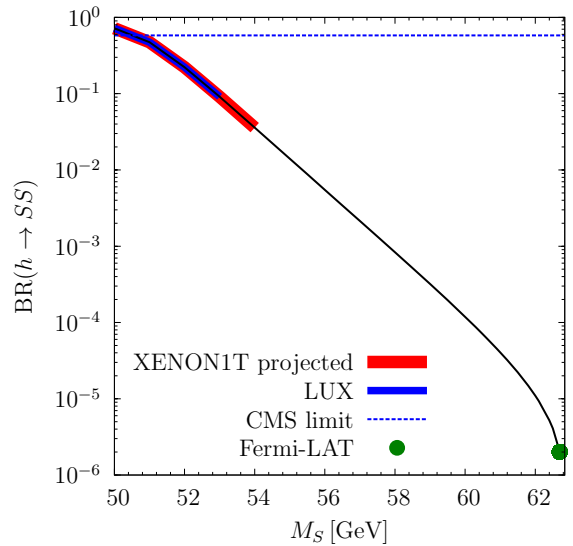}
 \caption{Invisible branching ratio, BR$(h \to S S)$, as a function of $M_S$. The blue dashed line corresponds to the CMS upper limit~\cite{Chatrchyan:2014tja}. 
 The blue solid part of the curve is excluded by the LUX experiment~\cite{Akerib:2013tjd}. The red line shows the exclusion potential of XENON1T~\cite{Aprile:2012zx}. The limit on $b \bar{b}$ from Fermi-LAT~\cite{Ackermann:2015zua} (green dot) constrains the region around the resonance and sets a lower limit in the invisible branching ratio. \label{fig:invisibledecay}} 
\end{figure}

One important comment on gamma line visibility is that beyond the $W^+W^-$ mass threshold the final state radiation from the $W$-bosons is no longer suppressed. The corresponding energy endpoint is at $E_\gamma^\text{max} = M_S \left(1 - M_W^2/M_S^2 \right)$. This leads to the conclusion that with an energy resolution between $5\%$ and $10\%$ one can only observe a line if $M_S \lesssim \unit[250]{GeV}$. Otherwise the line is smeared into the final state radiation from the $W$-bosons. This region will be investigated in a future publication.

Finally, using all the constraints from relic density, direct detection and indirect detection we show in Fig.~\ref{fig:invisibledecay} the predicted branching ratio for the invisible decay of the Standard Model Higgs. In blue we show the current LHC bound from CMS~\cite{Chatrchyan:2014tja}. Notice that the direct detection bound from LUX rules out the region where ${\rm{BR}} (h \to SS) > 10^{-1}$.  
This means that if the invisible decay width of the Higgs is large this model can be ruled out. Notice that there 
is a chance to test this model at LHC or future colliders if the invisible decay is not too small.

%%%%%%%%%%%%%%%%%%%%%%%%%%%%%%%%%%%%%%%%%%%%%%%%%%%%%%%%%%%%%%%%%%%%%%%%%%%%%%%
\section{Summary}
%%%%%%%%%%%%%%%%%%%%%%%%%%%%%%%%%%%%%%%%%%%%%%%%%%%%%%%%%%%%%%%%%%%%%%%%%%%%%%%
We have revisited the simplest model for scalar dark matter discussing all constraints coming from relic density, direct and indirect detection experiments, and the LHC. We have showed the allowed parameter space and discussed the possibilities to test this simple model, emphasizing the complementarity of all dark matter experiments. Additionally, we discussed the predictions for the annihilation into gamma rays in agreement with all constraints. 

For the first time we demonstrated that the gamma line can actually be observed in this model. This is due to the fact that the final state radiation cross section is suppressed by the small bottom Yukawa coupling.  Therefore, if the low mass version of this model is realized in nature one can clearly observe a gamma line in near future at gamma ray telescopes such as Fermi-LAT~\cite{Ackermann:2015lka} and the GAMMA-400~\cite{Topchiev:2015wva} experiments. 

%%%%%%%%%%%%%%%%%%%%%%%%%%%%%%%%%%%%%%%%%%%%%%%%%%%%%%%%%%%%%%%%%%%%%%%%%%%%%%%
{\textit{Acknowledgments}}:
The work of P.F.P.\ is partially funded by the Gordon and Betty Moore Foundation through Grant 776 
to the Caltech Moore Center for Theoretical Cosmology and Physics and Walter Burke Institute for Theoretical Physics, Caltech, Pasadena CA.
P.F.P.\ thanks the theory group at Caltech for hospitality at the end of this project.

%%%%%%%%%%%%%%%%%%%%%%%%%%%%%%%%%%%%%%%%%%%%%%%%%%%%%%%%%%%%%%%%%%%%%%%%%%%%%%%
%%%%%%%%%%%%%%%%%%%%%%%%%%%%%%%%%%%%%%%%%%%%%%%%%%%%%%%%%%%%%%%%%%%%%%%%%%%%%%%

%%%%%%%%%%%%%%%%%%%%%%%%%%%%%%%%%%%%%%%%%%%%%%%%%%%%%%%%%%%%%%%%%%%%%%%%%%%%%%%
%
%%%%%%%%%%%%%%%%%%%%%%%%%%%%%%%%%%%%%%%%%%%%%%%%%%%%%%%%%%%%%%%%%%%%%%%%%%%%%%%

%%%%%%%%%%%%%%%%%%%%%%%%%%%%%%%%%%%%%%%%%%%%%%%%%%%%%%%%%%%%%%%%%%%%%%%%%%%%%%%

\end{document}